\documentclass[showpacs,prb,twocolumn,amsmath,amssymb,floatfix]{revtex4}
\usepackage{graphicx}
\usepackage{dcolumn}
\usepackage{bm}
\usepackage{multirow}
\usepackage{subfigure}

\usepackage{color}

\newcommand{\Ecnl}{E_{\rm c}^{\rm nl}} 
 
\begin{document} 
 
\title{Adsorption structures of phenol on the Si(001)-(2$\times$1) surface calculated using density functional theory} 
\author{Karen Johnston} 
\affiliation{Max Planck Institute for Polymer Research, P.O.Box 3148,  
D 55021 Mainz, Germany} 
\author{Andris Gulans} 
\affiliation{COMP/Department of Applied Physics, Aalto University School of Science and Technology, P.O. Box 11100, FI-00076 AALTO, Finland}  
\author{Tuukka Verho} 
\affiliation{COMP/Department of Applied Physics, Aalto University School of Science and Technology, P.O. Box 11100, FI-00076 AALTO, Finland} 
\author{Martti J. Puska} 
\affiliation{COMP/Department of Applied Physics, Aalto University School of Science and Technology, P.O. Box 11100, FI-00076 AALTO, Finland} 
 
\begin{abstract} 
 
Several dissociated and two non-dissociated adsorption 
structures of the phenol molecule on the Si(001)-(2$\times$1) surface 
are studied using density functional theory with various exchange and 
correlation functionals.  The relaxed structures and adsorption 
energies are obtained and it is found that the dissociated structures 
are energetically more favourable than the non-dissociated structures. 
However, the ground state energies alone do not determine 
which structure is obtained experimentally.  To elucidate the 
situation core level shift spectra for Si $2p$ and C $1s$ 
states are simulated and compared with experimentally measured spectra.   
Several transition barriers were calculated in order to determine  
which adsorption structures are kinetically accessible.   
Based on these results we conclude that the molecule 
undergoes the dissociation of two hydrogen atoms on adsorption.   
 
\end{abstract} 
\maketitle 
\renewcommand{\baselinestretch}{2} 
\section{Introduction} 
 
Adsorption of organic molecules on semiconducting surfaces provides a 
potential way to produce smaller transistors 
\cite{Bent2002a,Zhou2006a,Zhou2006b}.  While there have been many 
studies concerning adsorption of benzene and related molecules on 
semiconductors there are surprisingly few studies concerning phenol 
adsorption.  The OH group gives rise to dissociative reaction 
possibilities, in addition to the non-dissociative adsorption observed 
for benzene.   
 
Casaletto {\it et al.} \cite{Casaletto2005a} studied phenol adsorption 
on silicon using photoemission spectroscopy.  Based on measurements of  
core level shifts (CLS) of the surface Si $2p$ states and C $1s$ states 
they concluded that phenol undergoes dissociative adsorption at room 
temperature and that in the adsorbed state the phenyl ring is bonded 
to the surface via the O atom (see structure D in 
Fig.~\ref{fig:structures}).  
However, since the structure cannot be directly observed it 
is possible that other structures could also fit the data.    
 
The theoretical study by Carbone {\it et al.} \cite{Carbone2007a} 
focused on possible adsorption sites for the non-dissociated 
butterfly structure (structure A in Fig.~\ref{fig:structures})  
and the above-mentioned dissociative structure. 
They looked at possible reaction paths between the structures and  
found that the conversion barrier is of the same order as the room  
temperature thermodynamic energy.  Other adsorption structures were  
not considered.    
 
In this article we report density functional calculations of 
a variety of phenol adsorption structures, which includes structures 
that were not discussed in Refs.~\onlinecite{Casaletto2005a,Carbone2007a}. 
In particular, we show that those structures should be included in the  
analysis.  The paper is organised as  
follows.  In section~\ref{sec:method} we describe the methodology, in 
section~\ref{sec:results} we present the structural data and 
adsorption energies of the various structures and calculate the core 
level shifts of the energetically most favored structures.   
The transition barriers to several adsorption states were calculated  
to determine which states are kinetically accessible.   
The results are discussed and summarised in section~\ref{sec:summary}.     
 
\section{Method} 
\label{sec:method} 
 
Adsorption energies calculated using density functional theory are 
known to depend remarkably on the exchange-correlation (XC) functional 
\cite{Hammer1999a}.  In order to understand this dependence more fully, 
we perform calculations using the generalized gradient functionals 
(GGA), PW91 \cite{Perdew1991a,Perdew1996b}, PBE 
\cite{Perdew1996a,Perdew1997a,Perdew1998a} and revPBE 
\cite{Perdew1996a,Zhang1998a,Perdew1998c}, the three-parameter hybrid  
functional B3LYP \cite{Becke1988a,Lee1988a} and the van der Waals density functional (vdW-DF)  
that includes non-local correlation to describe van der Waals  
interactions \cite{Dion2004a}.  
 
The PW91, PBE and revPBE calculations are performed using the 
Vienna {\it ab-initio} simulation package (VASP) 
\cite{Kresse1993a,Kresse1996b}, which is based on density functional 
theory (DFT) and uses a plane-wave basis set.  In the plane-wave 
calculations the core states are represented using the 
projector-augmented wave (PAW) method \cite{Blochl1994a,Kresse1999a} 
and the plane-wave cutoff energy is 400~eV. 
 
The B3LYP calculations are performed using an all-electron approach 
implemented in the LCAO (linear combination of atomic orbitals) code, 
CRYSTAL \cite{Crystal}.  We use the valence triple-$\zeta$ plus  
polarisation basis set for all atoms. The basis set for silicon was 
obtained by optimisation of the total energy of bulk silicon in 
Ref.~\onlinecite{BasisSi}. The basis sets for the remaining atoms acquired 
from Ref.~\onlinecite{TZVP} were orginally developed for atoms and 
molecules. In the present work, they are adapted for periodic 
calculations by increasing the exponent of the outermost p-type shell 
of C atom from 0.0892605 to 0.13.  This shrinks computational expenses 
and helps to avoid numerical problems such as quasi-linear dependence 
of basis functions \cite{Crystal}.    
 
In order to estimate the basis set superposition error (BSSE) in  
LCAO adsorption calculations, the counterpoise correction  
\cite{Counterpoise} is used. Its magnitude ranges from 0.24~eV to  
0.44~eV for structures A--F and from 0.60~eV to 0.75~eV for  
structures G and H shown in Fig.~\ref{fig:structures}.  This is 
a substantial correction and,  
therefore, the adsorption energies in the present paper always  
include it.  Nevertheless, even with the BSSE correction, a  
low quality basis set can yield energies far from the complete basis  
limit so we checked the performance of the basis set by comparing  
the PBE adsorption energies obtained with the LCAO and plane-wave  
approaches.  As shown in Table~\ref{tab:Eads}, the two methods give  
very similar results in all cases, except for structure H, where the  
difference in the adsorption energies is 0.26~eV.   
The BSSE introduces artifacts in the interaction 
between neighbouring phenyl-rings that strongly   
affects the energy and, due to the flexibility of the Si--O--C bonds, the  
molecule moves away from the true energy minimum.   
This problem cannot be resolved by a perturbative correction, however,  
if the basis set is expanded the relaxed structure changes and the LCAO  
adsorption energy starts to approach the plane-wave result.   
Consequently, this single case with a  
moderately large discrepancy is understood and the match between the  
LCAO and the plane-wave basis results for the other structures is  
good.  With this in mind, we conclude that our LCAO results  
are reliable.    
 
The van der Waals corrections to the adsorption energies 
are calculated using the real-space approach described in Ref. 
\onlinecite{Gulans2009a} combined with the multi-centre 
integration method \cite{Becke1992a}.   
Following Ref.\onlinecite{Dion2004a}, we use the revPBE functional 
\cite{Zhang1998a} to describe the exchange.  The total vdW-DF 
energy is calculated non-self-consistently as a post-GGA correction 
and is given by  
\begin{equation} 
\label{eqn:vdW-DF} 
E^{\textrm{vdW-DF}}=E^\textrm{revPBE}+E^\textrm{LDA}_{c}-E^\textrm{PBE}_{c}+\Ecnl, 
\end{equation}   
where $E^{\rm revPBE}$ is the total energy obtained in a 
self-consistent calculation with the revPBE XC 
functional. The next two terms, $E^\textrm{LDA}_{c}$ and $E^\textrm{PBE}_{c}$, are the LDA \cite{Perdew1992a} and PBE 
correlation energies, respectively. Their difference is calculated non-self-consistently using the PAW formalism. Finally, $\Ecnl$ is the non-local correlation term, which is evaluated using pseudo densities with the partial core correction.  
 
 
In all calculations we use the equilibrium Si lattice constant $a_0$, 
which is 5.47--5.49~{\AA} depending on XC functional used. 
The Si(001) surface is represented by nine atomic layers of Si atoms with the top 
side (2$\times$1)-reconstructed and the bottom layer Si atoms 
fixed in ideal lattice positions with their dangling bonds 
passivated by H atoms.  The positions 
of the Si atoms on the bottom layer and the passivating H atoms 
are held fixed.  In our VASP calculations, the supercell size for 
the 0.5~monolayer (ML) coverage\footnote{1~ML is defined here as 
1 molecule per Si dimer} is $\sqrt{2}a_0\times\sqrt{2}a_0\times 
6 a_0$, which contains a vacuum layer with a height of 
approximately 22~{\AA}.   
To check that there is no effect due to an 
artificial electric field, caused by using an asymmetric slab, 
structure D was recalculated with thicker vacuum layers.  The change 
in the total energy on going from $\sim$22~{\AA} to 
$\sim$38~{\AA} was only 0.004~eV.   
The CRYSTAL code employs the periodic boundary conditions only along 
the surface directions and hence the calculation does not include an 
artificial electric field in the vacuum.   
The Brillouin zone is sampled using a Monkhorst-Pack 
mesh of 4$\times$4$\times$1 k-points. The ionic relaxations are 
stopped when the maximum force on the ions is below 10~meV/{\AA}.    
 
To determine the transition states and barriers for the structural changes  
we use the adaptive nudged elastic band (ANEB) method  
\cite{Maragakis2002a}.   This calculation is essentially a search for
the saddle points of the potential energy surface and all obtained
transition states satisfy the same maximum force criterion as above.    
The total energy and its gradients are evaluated using the PBE XC 
functional.  
Due to the computational expense of these calculations we use thinner silicon  
slabs consisting of 5 layers of Si atoms and the Monkhorst-Pack mesh of  
2$\times$2$\times$1 $k$-points. This simplification results in 
changes in adsorption energies that are less than 0.07~eV.  
 
For a semi-quantitative analysis of reactions rate constants 
 we use the Arrhenius equation for the reaction rate constant 
\begin{equation} 
\label{eq:arrhenius} 
k = A \exp \left[-\frac{\Delta E}{k_{\rm B}T}\right]. 
\end{equation} 
$\Delta E$ is the energy barrier obtained from the ANEB calculations.  
In our estimates we use the temperature $T=293$~K, which is consistent with the  
experimental conditions in Ref.~\onlinecite{Casaletto2005a}.   
The pre-exponential factor $A$ is related to atomic vibrations and is assumed  
to be a constant.  We have used a value of $10^{12}$~s$^{-1}$, which is typical 
for reactions on surfaces.   
There are common inexpensive ways of estimating $A$, for example, using 
harmonic transition state theory, however, we have chosen to use a constant  
pre-exponential factor, since the right hand side of Eq.~\ref{eq:arrhenius} is much  
less sensitive to variations in $A$ than in $\Delta E$.

\section{Results and Discussion} 
\label{sec:results} 
 
\subsection{Structural data}   
 
\begin{figure*}[ht!]
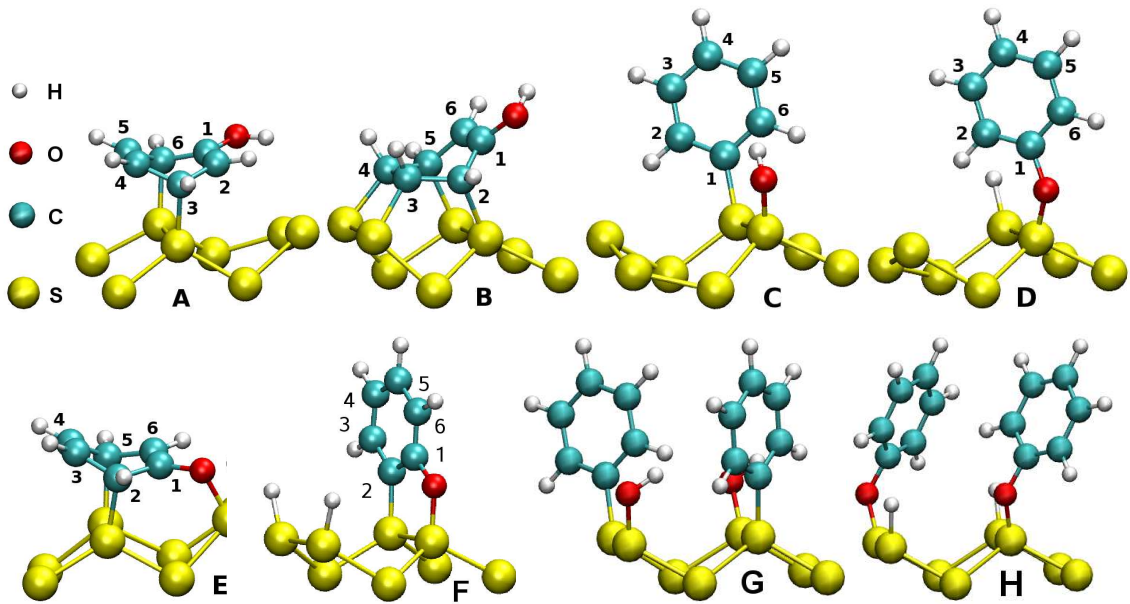
 
\begin{center} 
\includegraphics[height=3.0cm]{Fig1key.epsi} 
\includegraphics[height=2.5cm]{Fig1a.epsi} 
\includegraphics[height=3.0cm]{Fig1b.epsi} 
\includegraphics[height=4.0cm]{Fig1c.epsi} 
\includegraphics[height=4.0cm]{Fig1d.epsi} 
\end{center} 
\begin{center} 
\includegraphics[height=2.5cm]{Fig1e.epsi} 
\includegraphics[height=3.5cm]{Fig1f.epsi} 
\includegraphics[height=3.5cm]{Fig1g.epsi} 
\includegraphics[height=3.5cm]{Fig1h.epsi} 
\end{center} 
\caption{\label{fig:structures} (Color online) Locally stable adsorption 
structures for phenol on the Si(001)-(2$\times$1) surface. } 
\end{figure*}

The various adsorption structures are shown in 
Fig.~\ref{fig:structures}.  Structures A--F have a coverage of  
0.5~ML, whereas G and H correspond to C and D at a  
higher coverage of 1.0~ML.  The position of the OH group and/or 
dissociated H atoms corresponds to the energetically favorable 
position for each morphology.   
The discussion of energy barriers in subsection~\ref{sec:barriers} 
involves structures A', D', E' and F', which are not shown in 
Fig.~\ref{fig:structures}.  They are similar to structures A, D, E 
and F, respectively, but with the phenol molecule or with dissociated 
H atoms bonded to different sites.  
For instance, a phenol molecule in structure A is bound to one Si 
dimer, whereas in structure A' it is bound to Si atoms on adjacent dimers.   
 
Structures A and B are non-dissociated and are similar to the 
butterfly (BF) and tight-bridge (TB) structures of benzene 
on the Si(001)-(2$\times$1) 
surface\cite{Taguchi1991a,Wolkow1998a}.  In both structures A and B,  
the OH group remains bonded to a carbon atom.    
In structure C the OH group is dissociated from the phenyl ring, 
and the OH group and the phenyl ring are bonded to Si atoms on 
the same dimer.  
In structure D the O--H bond is broken and the C$_6$H$_5$O$^-$ radical  
and the dissociated H atom bond to Si atoms on the same dimer.   
Structure E is similar to A, but the O--H bond is broken and the  
O and H atoms are bonded to the Si dimer neighbouring the  
phenyl ring. The O--Si bond is slightly stretched compared to  
structures C and D due to the structural constraints.  
In structure F the O--H bond and the neighbouring C--H bond are 
broken and the two dissociated H atoms are adsorbed on the other Si 
dimer.  Similarly to structure E, the structural constraints result in 
a slightly longer Si--O bond compared to the free Si--O bond in  
structure C and D.  Structures G and H correspond to C and D but with 1.0~ML  
coverage.  The nearest neighbour phenyl rings are orthogonal with respect  
to each other.  Parallel orientation is energetically less favorable.

\subsection{Adsorption energies}   
\label{sec:energies} 
\subsubsection{Neglecting van der Waals interactions} 
 
The phenol adsorption energy is defined as  
\begin{equation} 
E_{\rm ads}=E_{\rm mol}+E_{\rm slab}-E_{\rm tot}, 
\end{equation} 
where $E_{\rm mol}$, $E_{\rm slab}$ are the total energies of the 
separate phenol molecule and the Si slab, respectively, and $E_{\rm 
  tot}$ is the total energy of the phenol molecule adsorbed on the 
Si-slab. The results obtained using the various XC 
functionals for the different structures are shown in 
Table~\ref{tab:Eads}.   
\begin{table}[ht!] 
\caption{\label{tab:Eads} Phenol adsorption energies (eV) on the 
  Si(001)-($2\times1$) per surface unit cell for structures 
  A-H. Results of standard DFT calculations with three different GGA 
  functionals as well as those of the B3LYP hybrid functional and 
  vdW-DF functional calculation with the revPBE exchange functional 
  are shown.}  
\begin{ruledtabular} 
\begin{tabular}{c|rrrr|r|r} 
      & \multicolumn{4}{c|}{Standard GGA} &Hybrid& Non-local \\ 
Structure & PW91 & PBE  &revPBE& PBE\footnotemark[1] & B3LYP\footnotemark[1] & vdW-DF\\ \hline 
A & 1.06 & 1.01 & 0.57 & 0.97 & 0.66 &  1.26 \\ 
B & 1.27 & 1.24 & 0.75 & 1.26 & 0.80 &  1.23 \\ 
C & 2.96 & 2.88 & 2.65 & 2.78 & 3.00 &  3.40 \\ 
D & 2.38 & 2.30 & 2.09 & 2.32 & 2.58 &  2.82 \\ 
E & 2.61 & 2.52 & 2.04 & 2.47 & 2.41 &  2.91 \\ 
F & 3.89 & 3.78 & 3.20 & 3.81 & 4.29 &  4.12 \\ \hline 
G & 5.85 & 5.68 & 5.03 & 5.18 & 5.52 &  7.16 \\ 
H & 4.82 & 4.64 & 4.13 & 4.40 & 4.92 &  6.07 \\ 
\end{tabular} 
\end{ruledtabular} 
\footnotetext[1]{LCAO} 
\end{table} 
 
The choice of the XC functional is known to 
influence adsorption energies significantly \cite{Hammer1999a}. 
PW91 and PBE have been found to give significantly higher 
adsorption energies than revPBE.  For 
the non-dissociated phenol structures A and B, the PBE 
adsorption energies are almost twice as large as those obtained with  
revPBE functional.  For the remaining 0.5 ML structures, the PW91 and 
PBE energies are   
larger than the revPBE energies by 0.29~eV--0.69~eV.  Despite 
these quantitative differences, the energetic ordering for both 
GGA functionals is mostly the same.  The dissociated structures are 
significantly lower in energy than the non-dissociated 
structures. This trend has been found previously for benzene 
\cite{Nunzi2007a} and for chloro- and dichlorobenzene 
\cite{Naumkin2003b}.

For a 0.5 ML coverage, the energetically most favorable structure is 
F, which is by 0.8--0.9~eV lower in energy than the next lowest energy 
structure C. However, according to the revPBE calculations D 
is 0.15~eV more favorable than E, whereas according to PW91 
and PBE calculations E is 0.22-0.23~eV more favorable than D.   
Carbone {\it et al.} \cite{Carbone2007a} used first-principles 
calculations with the BLYP functional \cite{Becke1988a,Lee1988a} to 
study structures A and D on various adsorption sites and found the 
adsorption energies of 0.55~eV and 2.56~eV, respectively.  This is in 
good agreement with our B3LYP results and the small differences are 
presumably due to the differences between BLYP and B3LYP functionals 
and the different coverages of $\frac{1}{6}$~ML and 0.5~ML used in 
their and the present calculations, respectively.  
 
The gain in energy due to the deposition is largely determined by the 
number of phenol molecules attached to the surface and it is limited 
by the available area. In structures E and F the maximum coverage has 
been reached already, while for structures C and D the number of 
adsorbed molecules can be doubled to form structures G and H.  Such  
an increase in coverage 
results in a gain of energy per surface unit, making G energetically  
favorable for all considered functionals.  
 
\subsubsection{Including van der Waals interactions} 
 
As shown in Table~\ref{tab:Eads}, van der Waals forces make the adsorption  
energies of A and B comparable, in contrast to the standard GGA results  
that predict the B to be more energetically favorable than A.  This is in  
agreement with the recent results for benzene on the same silicon  
surface\cite{Johnston2008a}.  
Despite qualitative similarities, the present adsorption energies of phenol  
and those of benzene in Ref.~\onlinecite{Johnston2008a} differ by $\sim0.45$~eV.   
Such a difference cannot be explained by a substitution of an H atom with  
the OH group.  We address this problem by repeating the calculations of  
Ref.~\onlinecite{Johnston2008a} within the present calculation scheme. The  
adsorption energies shown in Table~\ref{tab:cores} imply that the  
differences are of a numerical nature in evaluating the vdW correction.   
An analysis of the electron densities used by Johnston {\it et al.} in  
their calculations, reveals that the source of the discrepancy is the term  
$E^{\rm LDA}_{c}-E^{\rm PBE}_{c}$, which was calculated using the 
non-linear core correction, while in the present paper the PAW 
formalism is used.  Since the latter one is an all-electron method, we believe 
that the present results are more reliable.   
 
The reversal in the ordering is also observed for structures D and E.  
However, since there is a disagreement among the GGA functionals concerning  
the relative stability of these geometries we cannot draw any firm conclusions.  
 
Another interesting observation is that the high-coverage structures G 
and H have a higher adsorption energy per molecule than that of the  
corresponding low coverage structures C and D when the van der Waals  
interaction is included.  
The distance between the centres of neighbouring phenyl rings in 
structures G and H is $\sim$5.5~\AA, which is just slightly larger 
than the equilibrium distance between the molecules in an isolated 
benzene dimer in the T-shape and slip parallel configurations 
\cite{Puzder2006a}.  Due to such a geometrical layout on the 
Si(001)-($2\times1$) surface, the interaction of phenol molecules is 
attractive.   
 
In all cases the magnitude of the correction due to the van der Waals 
interaction (the last three terms in Eq.(\ref{eqn:vdW-DF})) is of the 
order of 0.48--1.07~eV per adsorbed molecule.   
The corrections are both large and scattered meaning that, in 
general, the correction may heavily influence the predictions for 
this kind of adsorbates.    
 
\begin{table}[ht!] 
\caption{\label{tab:cores} Benzene adsorption energies (eV) on the  
Si(001)-(2$\times$1) surface for structures BF and TB. The benzene  
coverage is 0.5~ML. The adsorption energies of  
Ref.~\onlinecite{Johnston2008a} are shown in parentheses.}  
\begin{ruledtabular} 
\begin{tabular}{c|cc|cc} 
 & \multicolumn{2}{c|}{Benzene} & \multicolumn{2}{c}{Phenol} \\ 
 Method & BF & TB & A & B\\ 
\hline 
PW91   & 1.00(0.99)& 1.25(1.24) &1.06 & 1.27 \\ 
PBE    & 0.96(0.92)& 1.24(1.19) &0.97 & 1.26\\ 
revPBE & 0.53(0.48)& 0.72(0.66) &0.57 & 0.75 \\ 
vdW-DF & 1.13(0.82)& 1.15(0.77) & 1.26& 1.23\\ 
\end{tabular} 
\end{ruledtabular} 
\end{table}

\subsection{Core level shifts}  
 
At this stage we focus on the dissociated structures and neglect 
structures A and B.  So far, the results show that, for a coverage of 
0.5~ML, C and F are more stable than D and that E has an energy  
comparable to that of D.  At high phenol exposure, corresponding to  
increased coverage, structure G becomes the energetically favourable one.  
In this subsection we attempt to determine the experimentally observed 
structure by calculating the CLS spectra for each structure.   
 
The experimental study by Casaletto {\it et al.}  
\cite{Casaletto2005a} used X-ray photoemission spectroscopy (XPS) to  
measure the CLSs for the Si $2p$, C $1s$, and O $1s$ core states.  In the  
present work, C $1s$ and Si $2p$ CLSs for structures C--F and H (the  
omission of G is explained below) are calculated and compared with  
experimental data.  In calculating the core-level shifts, we use the  
method described in Ref.~\onlinecite{Yazyev2006a}, where a pseudopotential  
for an atom core with a hole is constructed. Then, in the supercell 
calculation, an electron is removed from the system and a homogeneous  
background charge is applied to keep the system neutral.  We use VASP  
with the PW91 functional for these calculations.  To calculate the  
relative C $1s$ CLSs for phenol on the Si(001)-(2$\times$1) we use a  
nine atomic layer Si slab with the H-passivation on the bottom  
surface as described above.  The Si 2$p$ CLSs were found to be much  
more sensitive to the slab thickness than the structural properties  
and energies are. Thus, to obtain converged Si 2$p$ CLSs we used  
slabs with 17 layers of Si atoms.   
 
\subsubsection{C $1s$ core level shifts}  
 
Table~\ref{tab:cls-C} shows the C $1s$ CLSs for structures C--F and H.   
It is clear from the results for D and H that the coverage does not affect  
the C $1s$ CLS, therefore, we have not calculated the CLSs for G as the  
results would be equal to those for C.   
For each structure, the average core level binding energy of 
the carbon atoms $sp^2$-bonded to two other carbon atoms and to one hydrogen  
atom is taken as the reference energy.  
\begin{table}[ht!] 
\caption{\label{tab:cls-C} C $1s$ core level shifts (eV) for phenol adsorbed 
on the Si(001)-(2$\times$1) surface. The different adsorption structures, 
C--F and H, and the labelling of carbon atoms are shown in  
Fig.~\ref{fig:structures}. For each structure the  
reference energy is the core level binding energy averaged over the 
$sp^2$-bonded, benzene-like carbon atoms.} 
\begin{center} 
\begin{tabular}{l|rrrrr|r}\hline\hline 
   & C      &  D    &  E    &  F  &  H  & Model, Ref.~\onlinecite{Casaletto2005a} \\ \hline 
C1 & $-0.35 $ & $ 1.70$ & $ 1.66$ & $ 1.62$ & $ 1.71$ & 1.5 \\ 
C2 & $-0.10 $ & $-0.12$ & $ 0.41$ & $-0.42$ & $-0.09$ & 0 \\ 
C3 & $ 0.04 $ & $ 0.14$ & $ 0.14$ & $ 0.05$ & $ 0.09$ & 0 \\ 
C4 & $ 0.01 $ & $-0.09$ & $ 0.07$ & $-0.09$ & $-0.02$ & 0 \\ 
C5 & $ 0.10 $ & $ 0.14$ & $ 0.47$ & $ 0.06$ & $ 0.11$ & 0 \\ 
C6 & $-0.04 $ & $-0.08$ & $-0.21$ & $-0.02$ & $-0.08$ & 0 \\ 
\hline 
\end{tabular} 
\end{center} 
\end{table} 
Experiments by Casaletto \textit{et al.} \cite{Casaletto2005a} 
showed two peaks, which 
were attributed to the carbon atoms in the phenyl-ring and the carbon atom 
bonded to the oxygen atom.  Based on the magnitude of the shift and the 
ratio of 1:5 of the integrated intensity, Casaletto {\it et al.}  
concluded that the structure D was observed.    
According to our calculations, the spectra of structures D--F and H all have 
one CLS around 1.6--1.7~eV with respect to the reference energy. For  
structures D and H the remaining five CLSs are positioned in the narrow  
range, $-0.12$ to $+0.14$~eV. Such a compact grouping can be explained by  
almost equivalent nearest neighbour surroundings of the carbon atoms labelled with  
numbers 2--6 in Fig.~\ref{fig:structures}.  This is not the case for  
structures E and F, in which some of the carbon atoms are bonded to silicon 
atoms and have different environments than the other carbon atoms.  
Consequently the spectra contain CLSs that are shifted from the  
reference energy by $0.4$--$0.5$~eV and $-0.4$~eV for E and F,  
respectively.   
 
\begin{figure}[ht!] 
\begin{center} 
\includegraphics[width=7cm,angle=0]{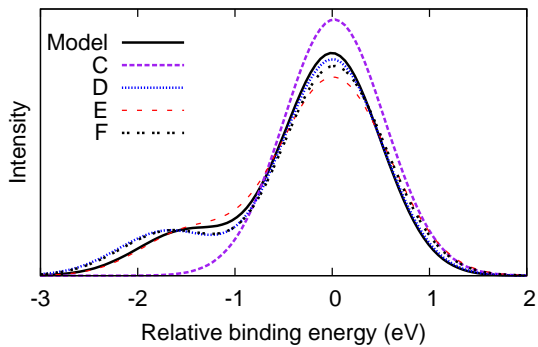} 
\caption{\label{fig:cls-C} (Color online) Simulated C $1s$ core level 
 binding energy spectra for structures C, D, E and F. The energy zero 
is the core level binding energy averaged over the 
$sp^2$-bonded, benzene-like carbon atoms. The model function is based on the  
measured spectrum in Ref.~\onlinecite{Casaletto2005a}}  
\end{center} 
\end{figure}

Although there are spectral features unique to each of structures  
C, D, E and F, they are not necessarily resolvable in experiment. 
In order to compare the calculated results directly to experiment,  
we plotted the core level shifts 
using Gaussian functions with the experimental FWHM of 1.0~eV. 
The curves are shown in Fig.~\ref{fig:cls-C} along with a model 
function, which was constructed to reproduce the line-shape 
analysis and experimental data in Ref.~\onlinecite{Casaletto2005a}.   
The figure shows clearly that the spectra of structures D (H), E and F 
qualitatively fit the model function.  However, since there is no visible 
shoulder for structure C (G) we can rule it out.   
 
\subsubsection{Si $2p$ core level shifts} 
 
Next, we consider whether it is possible to distinguish between  
structures D, E, F and H on the basis of the Si $2p$ surface core level  
shifts. To extract information about the Si $2p$ CLSs for the surfaces with 
adsorbed phenol it is first necessary to calculate the shifts for 
the clean surface.  The CLSs for the clean surface and for structures 
D, E, F and H are shown in Table~\ref{tab:cls-Si}.  For 
these calculations the silicon slab contains 17 atomic layers and the reference energy is  
taken to be the core level binding energy averaged 
over the bulk-like 13th--16th Si layers below the reconstructed surface.    
The Si 2$p$ CLSs with adsorbed phenol do not agree with the data by  
Casaletto {\it et al.} \cite{Casaletto2005a}.  This is most 
likely due to their assumption that only the CLSs of the Si 
atoms on the surface are affected by the phenol adsorption, whereas  
our calculations clearly show that the SCLSs of the 
subsurface atoms change significantly.  
 
\begin{table}[ht!] 
\caption{\label{tab:cls-Si} Relative Si $2p$ core level shifts (eV) for the 
  clean Si(001)-(2$\times$1) surface and for the surface after phenol 
adsorption. Adsorbate structures D, E and F with the 0.5~ML coverages  
and structure H with the 1~ML coverage are shown in Fig.~\ref{fig:structures}.  
The reference energy is the binding energy in the bulk environment.}  
\begin{center} 
\begin{tabular}{rl|r|rrr|r|r}\hline\hline 
Layer & Atom    & Clean   & D       & E       & F       & H       & Expt.~\cite{Casaletto2005a} \\ \hline 
    1 & Si up   & $-0.64$ & $-0.68$ & $-$     & $-$     & $-$     & $-0.523$ \\ 
    1 & Si down & $+0.01$ & $-0.02$ & $-$     & $-$     & $-$     & $+0.097$ \\ 
    1 & Si--C2  & $-$     & $-$     & $+0.06$ & $-0.12$ & $-$     & $-$ \\ 
    1 & Si--C5  & $-$     & $-$     & $+0.14$ & $-$     & $-$     & $-$ \\ 
    1 & Si--O   & $-$     & $+0.69$ & $+1.00$ & $+0.56$ & $+0.82$ & $+0.922$ \\ 
    1 & Si--O   & $-$     & $-$     & $-$     & $-$     & $+0.85$ & $-$ \\ 
    1 & Si--H   & $-$     & $-$     & $+0.08$ & $+0.18$ & $+0.03$ & $-$ \\ 
    1 & Si--H   & $-$     & $+0.08$ & $-$     & $+0.15$ & $+0.03$ & $+0.344$ \\ \hline 
 
    2 & Si      & $-0.10$ & $-0.21$ & $-0.24$ & $-0.07$ & $-0.17$ & $+0.224$ \\ 
    2 & Si      & $+0.09$ & $-0.18$ & $-0.11$ & $-0.07$ & $-0.16$ & $-0.232$ \\ 
    2 & Si      & $-$     & $-0.01$ & $-0.02$ & $-0.07$ & $-0.10$ & $-$ \\ 
    2 & Si      & $-$     & $ 0.00$ & $+0.15$ & $-0.07$ & $-0.09$ & $-$ \\ \hline 
 
    3 & Si      & $+0.34$ & $-0.21$ & $-0.22$ & $-0.19$ & $-0.22$ & $-$ \\ 
    3 & Si      & $-0.09$ & $-0.19$ & $-0.05$ & $-0.19$ & $-0.20$ & $-$ \\ 
    3 & Si      & $-$     & $+0.16$ & $+0.14$ & $+0.05$ & $+0.11$ & $-$ \\ 
    3 & Si      & $-$     & $+0.18$ & $+0.20$ & $+0.05$ & $+0.11$ & $-$ \\ \hline 
 
    4 & Si      & $-0.26$ & $-0.37$ & $-0.23$ & $-0.22$ & $-0.20$ & $-$ \\ 
    4 & Si      & $+0.23$ & $-0.09$ & $-0.13$ & $-0.18$ & $-0.20$ & $-$ \\ 
    4 & Si      & $-$     & $+0.07$ & $+0.05$ & $+0.01$ & $+0.01$ & $-$ \\ 
    4 & Si      & $-$     & $+0.08$ & $+0.07$ & $+0.06$ & $+0.01$ & $-$ \\ 
\hline 
\end{tabular} 
\end{center} 
\end{table} 
 
\begin{figure}[ht!]
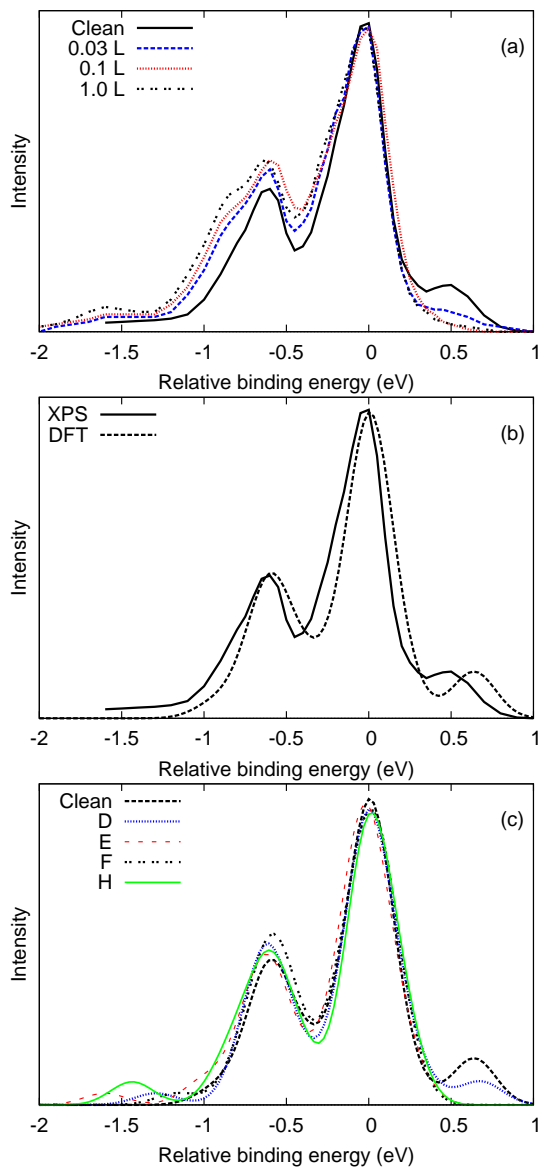
 
\begin{center} 
\includegraphics[width=7.0cm,angle=0]{Fig3a.epsi} \  
\includegraphics[width=7.0cm,angle=0]{Fig3b.epsi} \  
\includegraphics[width=7.0cm,angle=0]{Fig3c.epsi} 
\caption{\label{fig:Si2p-CLS} (Color online) Si $2p$ core level 
  shifts for the Si(001)-(2$\times$1) surface.  (a) The 
  XPS data from Ref.~\onlinecite{Casaletto2005a} for various phenol 
  concentrations.  (b) Fit of the DFT data to the experimental data 
  for the clean surface.  (c) DFT curves for structures D, E, F and H and the clean surface.  }  
\end{center} 
\end{figure}

Although the curve fitting to the experimental data is inaccurate 
we can still make use of the raw experimental data as shown in 
Fig.~\ref{fig:Si2p-CLS}.  The experimental curve for the clean surface 
has three distinct peaks. Due to the spin-orbit splitting of the $2p$ level 
two peaks with the intensity ratio of 1:2 and the separation of $s=0.602$~eV. 
\cite{Casaletto2005a} correspond to each different Si atom environment. 
The peak at 0.5~eV corresponds to the Si up-dimer atom and disappears 
as the coverage increases. With the increasing coverage, also the peak  
at $-0.6$~eV develops a shoulder at $-0.8$--$-1.0$~eV and a weak peak develops  
at around $-1.6$~eV.  The shoulder and the small peak probably belong to the same 
atoms as their separation is approximately equal to the spin-orbit 
splitting.   
 
We start by plotting the data for the clean surface and by comparing to 
experiment.  The CLS for each atom is chosen to be a sum of two Gaussian 
functions with the above-mentioned intensity ratio and energy splitting.  
The total intensity of the simulated spectruim is then equal to a sum of  
these split Gaussian functions, i.e.,  
\begin{equation} 
\begin{split} 
I(x) &= \sum_{i=1}^N \alpha^{L-1} \left\{2 e^{-(x+a_i+s)^2/2b^2} 
     + 4 e^{-(x+a_i)^2/2b^2}\right\} \\ 
     &+ N_{\rm bulk} \left\{2 e^{-(x+s)^2/2b^2} + 4 
     e^{-(x)^2/2b^2}\right\}. 
\end{split} 
\end{equation} 
Above, $a_i$ is the the core level shift for atom $i$.  $2b$ is the 
FWHM, for which we use the value of 0.26~eV. This is the average of  
the values for bulk and surface atoms used by Casaletto {\it et al.} 
\cite{Casaletto2005a}. 
$0<\alpha<1$ is an attenuation constant, which weakens the 
contribution from the subsurface layers and $L$ is the layer 
index, so that $L=1$ corresponds to the surface layer, $L=2$ to the 
subsurface atoms, etc.   
$N_{\rm bulk}$ and $\alpha$ are parameters chosen 
to fit the calculated clean surface spectrum to the experimental one. 
Using $\alpha=0.7$ and $N_{\rm bulk}=7$ reproduces well the main 
features of the experimental curve, as can be seen in 
Fig.~\ref{fig:Si2p-CLS}.  The simulated spectra for 
structures D, E, F and H in Fig.~\ref{fig:Si2p-CLS} are calculated using  
the same parameters as for the clean surface.  To analyse our results  
it is easiest to observe the changes in the spectra for the different 
structures with respect to the clean surface spectrum.   
 
In our fit the three main peaks are visible although somewhat 
shifted compared to the experimental data.  This is probably due to 
the reference value of the core level shifts not being equal to the 
true bulk value.  Nevertheless the fit is good enough to compare 
qualitatively, as shown in Fig.~\ref{fig:Si2p-CLS}.   
The most obvious change in the curve is the disappearance of the Si 
up-dimer peak, which implies that the surface is saturated and that no 
asymmetric dimers remain.  This saturation occurs for structures E, F and  
H but not for D.   
It is senseless to discuss which structure agrees best with the experimental  
Si $2p$ CLS spectrum, since, as explained before, we can make only a qualitative  
comparison.  It is clear that none of the three structures can be ruled out  
on the basis of the data provided by CLS spectroscopy.

\subsection{Reaction barriers}  
\label{sec:barriers} 
 
\begin{figure}[ht!] 
\begin{center} 
\includegraphics[width=9cm,angle=-90]{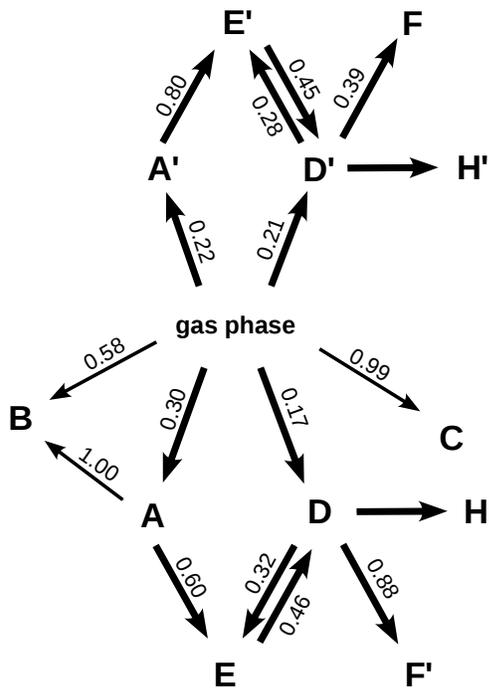} 
\caption{\label{fig:flowchart}   
Transition barriers (in eV) between the considered adsorption structures.  Thick arrows represent probable reactions, whereas thin arrows represent unlikely ones.}
\end{center} 
\end{figure}

So far we can conclude that any of structures D (H), E or F would 
be consistent with the experimental core level shift data.   
To determine which of the conformations are accessible at the room temperature 
(used for experimental observation) we have calculated the activation energies  
for different structural transformations of adsorbed phenol molecules as shown  
in Fig.~\ref{fig:flowchart}.   
 
\begin{figure}[ht!]
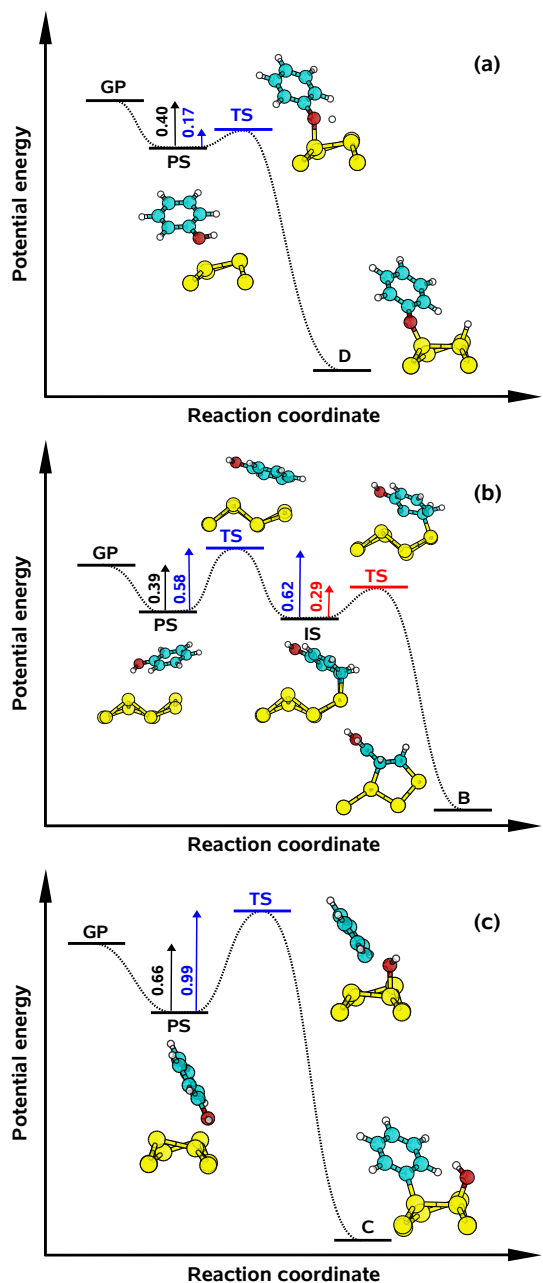
 
\subfigure{\includegraphics[width=5.5cm,angle=-90]{Fig5a.epsi}\label{fig:GP-D}}
\subfigure{\includegraphics[width=5.5cm,angle=-90]{Fig5b.epsi}\label{fig:GP-B}} 
\subfigure{\includegraphics[width=5.5cm,angle=-90]{Fig5c.epsi}\label{fig:GP-C}}
\caption{\label{fig:GP-ABCD}(Color online) A sketch of the minimum 
  energy path for molecule adsorption. The final states are 
  structures (a) D, (b) B and (c) C.  Gas-phase states, precursor 
  states, transition states and an intermediate, locally stable state
  are marked as GP, PS, TS and IS, respectively.  The numbers
  indicate the activation energies in eV. }    
\end{figure} 
The adsorption reactions starting from the gas phase always involve  
precursor states, which can be seen in Fig.~\ref{fig:GP-ABCD}.   
These precursor states are not discussed further as they only serve as 
initial traps where phenol molecules are bound non-covalently and 
weakly ($<0.7$~eV).  In the course of time, the molecules either 
detach from the surface or transform to one of structures A--D.   
The transition from the molecule in the gas phase to structure D is  
shown in Fig.~\ref{fig:GP-D}.  
After the molecule becomes trapped in the precursor state it faces a $0.17$~eV  
high energy barrier on its path to structure D.  On the other hand, the  
energy needed for returning to the gas phase is $0.40$~eV.  Inserting these  
energies into Eq.~\ref{eq:arrhenius} yields a $\sim10^4$ times greater reaction rate  
constant for transforming to structure D than for desorption.  Qualitatively the  
same picture is observed for the adsorption reaction with structure A as the final  
product.  
The opposite conclusion can be drawn for the formation of structures B and C.   
The energetics of these transitions are shown in Figs.~\ref{fig:GP-B}
and \ref{fig:GP-C}.  
The main difference in these curves compared to Fig.~\ref{fig:GP-D}  
is the height of the barrier that a molecule has to overcome in order to form  
a covalent bond.  For structures B and C this energy is noticeably higher than  
for desorption, therefore their formation is improbable.  
 
The flowchart in Fig.~\ref{fig:flowchart} does not contain reactions 
where structures E and F are acquired directly from the gas 
phase. These reactions require a formation of intermediate products 
such as structures A or D.  
For instance, structure F is produced by breaking two bonds: O--H and C--H.   
In this case the former bond is much easier to break than the latter one and  
we anticipate that in the first dissociation event the hydrogen atom 
splits off from the O atom.  This corresponds to the formation of a 
D-like structure.  Thus, we consider reaction D$\rightarrow$F rather 
than the adsorption  
of a molecule directly to structure F.  On closer inspection, reaction 
D$\rightarrow$F involves the diffusion of hydrogen atoms on the
surface.  
Our calculations show that the energy barrier for a hydrogen atom
  to move diagonally across the dimer row is of the 
order of $2.5$~eV, which is consistent with the findings of Bowler
{\it et al.}\cite{Bowler2000a}.  Hence, we disregard this particular
reaction in further discussion.  On the other hand, reactions 
involving primed structures, i.e. D$\rightarrow$F' and D'$\rightarrow$F, 
do not require such a diffusion and the only barrier to overcome is 
related to the cleavage of a C--H bond. 
 
The activation energies can be inserted into Eq.~\ref{eq:arrhenius} 
to calculate reaction rate constants.  The slowest 
transitions are A$\rightarrow$E and D$\rightarrow$F', for which 
$1/k=2\times10^{-2}$~s and $1/k=20$~min, respectively.
This shows that the formation of structure F' is slow, yet it cannot be 
ignored. 
 
\begin{figure}[ht!] 
\begin{center} 
\includegraphics[width=5.0cm,angle=-90]{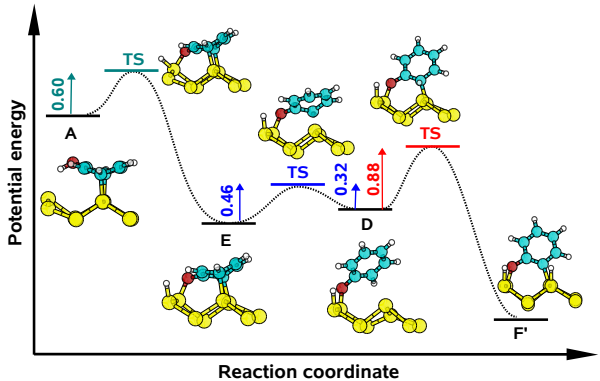} 
\caption{\label{fig:AE-DE-DF} (Color online)  
A sketch of the minimum energy path for reactions A$\rightarrow$E, 
E$\rightleftarrows$D and D$\rightarrow$F'.  Transition states are 
  labelled TS.  The numbers indicate the activation energies in eV.  }  
\end{center} 
\end{figure}

Consider a phenol molecule that approaches the surface with an 
orientation that leads to structures A' or D'. Then the molecule
undergoes a sequence of structural transformations
A'$\rightarrow$E'$\rightleftarrows$D'$\rightarrow$F, which is
qualitatively similar to the one sketched in 
Fig.~\ref{fig:AE-DE-DF}.  
However, in this case the dissociation of a 
C--H bond requires only 0.39~eV, which is a much lower energy than the
0.88~eV required for the reaction D$\rightarrow$F'.  
To explain this difference we notice that in structures F and F' the
molecule is bound to two Si atoms that are separated by
2.4~\AA~and 3.8~\AA , respectively.  The total energy of the former
configuration is lower by 0.35~eV, which indicates that due to the
variation of distance between the two Si atoms an additional strain is
exerted on the molecule in structure F'. The same geometry
considerations are valid for the two transition states, and the
difference in their deformation energies can be estimated by the same
number as above. This roughly corresponds to the difference in energy
barriers for reactions D$\rightarrow$F' and D'$\rightarrow$F.  
 
Molecules with structures F and F' are bound the most 
strongly by some margin over the other structures so they are the 
final products of the reactions that happen on the surface. 
However, the slow time scale of reaction D$\rightarrow$F' implies 
that there can be other faster processes that prevent the formation 
of structure F'.  For example, if the surface is exposed to a high 
phenol pressure and if we assume that the activation energy for 
D$\rightarrow$H is similar to the gas-phase$\rightarrow$D reaction 
then the surface will saturate to form structure H.  Obviously, our 
analysis is incapable of providing quantitative information about 
what pressures are required in order for this outcome to take place.   
Instead, we note that in the experimental results for lower phenol 
exposures the complete coverage of the surface is not 
reached\cite{Casaletto2005a} and under these conditions structure 
F', rather than H, will be obtained.   
For the reaction D'$\rightarrow$F the energy barrier is low and, 
consequently, the transition time is fast. In fact, the whole 
sequence gas-phase$\rightarrow$D($\rightleftarrows$E)$\rightarrow$F 
involves only fast reactions.  This means that other processes such as 
the formation of structure H' are extremely unlikely to interfere 
and at room temperature structure F will be abundant on the surface 
at any phenol deposition conditions. 
 
To our knowledge, activation energies of the considered reactions
have not been measured experimentally.  However, the quality of
the present calculations can be indirectly assessed by using
available data on the adsorption of benzene on
Si(001)-($2\times1$).  The calculated activation energy of the
A$\rightarrow$B reaction is 1.00~eV, which is in good agreement with
the experimentally measured barrier of 0.95~eV for the structural
transformation BF$\rightarrow$TB for benzene adsorbed on the
Si(001)$-$(2$\times$1) surface \cite{Lopinski1998a}.

\section{Conclusions} 
\label{sec:summary} 
 
Density functional theory calculations of the adsorption of phenol on the  
Si(001)-(2$\times$1) surface were performed.   
Regardless of the XC functional used, we found that the  
dissociated structures were energetically more stable than the  
non-dissociated ones.  The highest adsorption energy per phenol molecule,  
obtained for the structure with two dissociated hydrogen atoms (structure F),  
is 3.20--4.29~eV.  On the other hand, the highest energy per surface unit cell,  
obtained for the 1~ML coverage structure with one dissociated hydrogen atom  
(structure G), is 4.13--6.07~eV.  The large range of adsorption energies 
shows the strong dependence on the XC functional used.   
An important effect is observed when van der Waals interactions are included.   
Namely, similar to benzene, the relative stability of the structures is  
affected when the van der Waals interaction is included in the calculations.   
Furthermore, for a 1~ML coverage, van der Waals forces cause  
an attraction between neighbouring phenyl rings.  
 
C $1s$ and Si $2p$ CLS spectra for the dissociative structures were 
simulated and compared with the photoemission spectra in 
Ref.~\onlinecite{Casaletto2005a}.  Based on the comparison, we found that the 
structures with the cleaved OH group, C and G, do not fit the 
C $1s$ spectra obtained in experiment.   
The disappearance of the Si up-dimer peak from the experimental Si $2p$  
spectra suggests that the surface is fully saturated and thus we can rule out the structure D.   
The previous analysis of the C $1s$ CLS spectrum led to the conclusion that  
structure H (or D) was observed\cite{Casaletto2005a}.  However, we have shown  
that the remaining dissociative structures, E and F, have very similar  
C $1s$ and Si $2p$ CLS spectra to H and, therefore, they can not be distinguished from H  
using photoemission spectroscopy alone.   
 
From an analysis of reaction barriers we have shown that the 
activation energies for the formation of structures F and F' are 
0.39~eV and 0.88~eV, respectively. They are low enough that both 
reactions will occur at room temperature. However, the rate of 
formation of structure F' is slow and at high phenol pressure 
conditions it will be replaced by structure H.  The low barrier path 
to F suggests that this structure will be the most abundant.  
 
\begin{acknowledgments} 
We thank Denis Andrienko and Nico van der Vegt for  
critical reading of the manuscript and Risto Nieminen for valuable discussions.    
We acknowledge computational resources provided by the Finnish IT Center  
for Science (CSC) and by the Rechenzentrum (RZG) of the Max Planck Society. 
This research was supported by the Finnish Funding Agency for Technology 
and Innovation (TEKES), the Academy of Finland through its Centres of  
Excellence Program (2006--2011) and by the Multiscale Modelling Initiative 
of the Max Planck Society.  
 
\end{acknowledgments} 
 
\bibliography{proof} 
 
\end{document}